\documentclass[twocolumn,english,letterpaper]{revtex4}
\pdfoutput=1
\usepackage{times}
\usepackage[T1]{fontenc}
\usepackage[latin1]{inputenc}
\usepackage{geometry}
\usepackage{amsmath}
\usepackage{color}
\usepackage{graphicx}
\usepackage{amssymb}

\makeatletter
\usepackage{babel}
\makeatother
\begin{document}

\title{Effective-range description of a Bose gas under strong one- or two-dimensional
confinement}

\author{Pascal Naidon$^{a}$, Eite Tiesinga$^{a}$, William F. Mitchell$^{b}$,
and Paul S. Julienne$^{a}$}

\date{\today}

\affiliation{$^{a}$Atomic Physics Division, National Institute of Standards and
Technology, 100 Bureau Drive Stop 8423, Gaithersburg, Maryland 20899-8423,
USA}

\affiliation{$^{b}$Mathematical and Computational Sciences Division, National
Institute of Standards and Technology, 100 Bureau Drive Stop 8910,
Gaithersburg, Maryland 20899-8910, USA}

\begin{abstract}
We point out that theories describing $s$-wave collisions of bosonic
atoms confined in one- or two-dimensional geometries can be extended
to much tighter confinements than previously thought. This is achieved
by replacing the scattering length by an energy-dependent scattering
length which was already introduced for the calculation of energy
levels under 3D confinement. This replacement accurately predicts
the position of confinement-induced resonances in strongly confined
geometries.
\end{abstract}
\maketitle
Many experiments investigating the properties of cold atomic gases
and Bose-Einstein condensates are now performed in tightly confining
traps, such as tight optical lattices, leading to systems of reduced
dimensionality \cite{gorlitz2001,denschlag2002,richard2003,spielman:020702}.
There are many uses for such confinements. In spectroscopic measurements,
they eliminate unwanted Doppler and recoil effects \cite{ido2003,ludlow2006}.
They can also be used to create tunable analogs of condensed matter
systems, and give the possibility to investigate remarkable many-body
regimes in low dimensions such as the Tonks-Girardeau gas \cite{olshanii1998,petrov2000b,paredes2004,kinoshita2004}.
The theory of $s$-wave atomic collisions in strongly confined systems
has been established in Refs.~\cite{olshanii1998} and \cite{petrov2000a}
for 2D and 1D confinement, respectively. Both predict a confinement-induced
resonance of the effective 1D or 2D interaction strength. These predictions
rely on a description of the atomic interaction in terms of the scattering
length only. However, in 3D confined systems, it was shown that a
more refined description is needed for very tight confinement \cite{tiesinga2000,blume2002}.
Similarly, in 2D confined systems, numerical calculations in Ref.~\cite{bergeman2003a}
showed that the scattering length description of Ref.~\cite{olshanii1998}
may be insufficient. In this paper, we present an accurate analytical
description for scattering in 1D and 2D geometries based on the findings
of Refs.~\cite{blume2002,bolda2002}.

We consider a gas of bosonic atoms in an optical lattice and assume
that there is little tunelling between the lattice cells, so that
each cell is independent. The atoms in a cell are confined by a trapping
potential which will be assumed harmonic (which is true near the centre
of the cell). Let us consider a pair of atoms in such a cell. For
a harmonic potential, the centre-of-mass motion decouples from the
relative motion and the stationary Schr\"odinger equation for the
relative motion wave function $\psi(\vec{r})$ reads:\begin{equation}
\left[-\frac{\hbar^{2}}{2\mu}\nabla_{\mathbf{r}}^{2}+U(r)+V(\vec{r})\right]\psi(\vec{r})=E\psi(\vec{r}).\label{eq:Schrodinger}\end{equation}
Here, $\vec{r}=(x,y,z)$ is the relative coordinate with separation
$r$, $\mu$ the reduced mass, $U(r)$ the isotropic atom-atom interaction
potential, $V(\vec{r})$ the trapping potential, and $E$ the relative
energy.

For 2D confinement (\emph{tube} or \emph{wave guide} geometry), the
atoms are strongly confined in the $xy$ directions and (almost) free
to move in the $z$ direction, therefore we set:\[
V(\vec{r})\equiv V_{2D}(\vec{r})=\frac{1}{2}\mu\omega^{2}\rho^{2}\]
where $\rho=\left\Vert \vec{\rho}\right\Vert $ and $\vec{\rho}\,$
is the projection of $\vec{r}$ on the $xy$ plane. For 1D confinement
(\emph{pancake} geometry), the atoms are strongly confined in the
$z$ direction and (almost) free to move in the $xy$ directions:\[
V(\vec{r})\equiv V_{1D}(\vec{r})=\frac{1}{2}\mu\omega^{2}z^{2}\]
Here, $\omega$ is the trapping frequency at the centre of the cell
and we define $\sigma=\sqrt{\hbar/(2\mu\omega)}$ as the typical length
scale associated to the trap in the confined directions.

\begin{figure*}[t]
\includegraphics[scale=0.8]{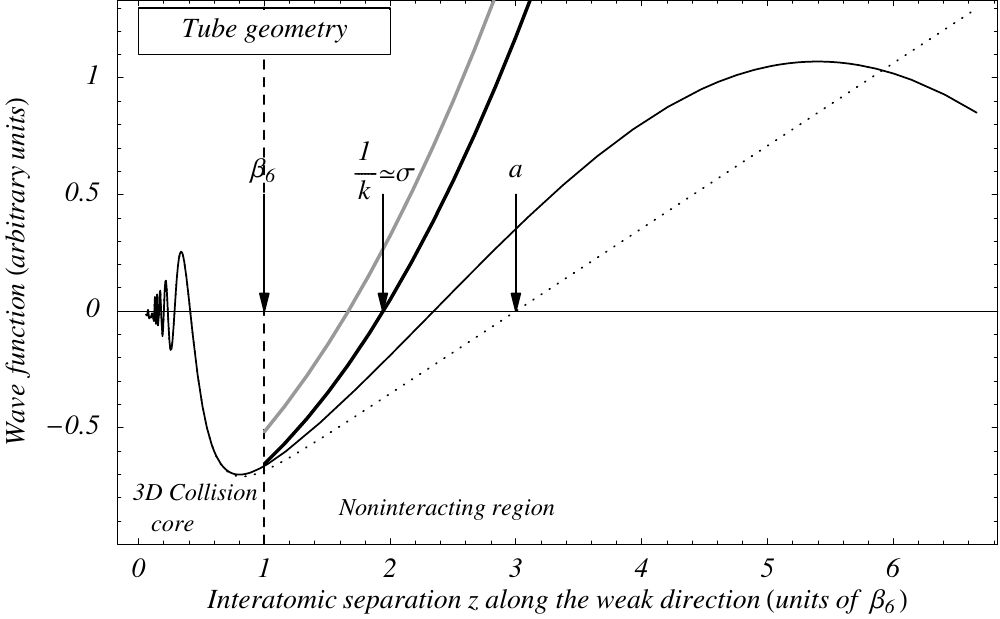}\hfill{}\includegraphics[scale=0.8]{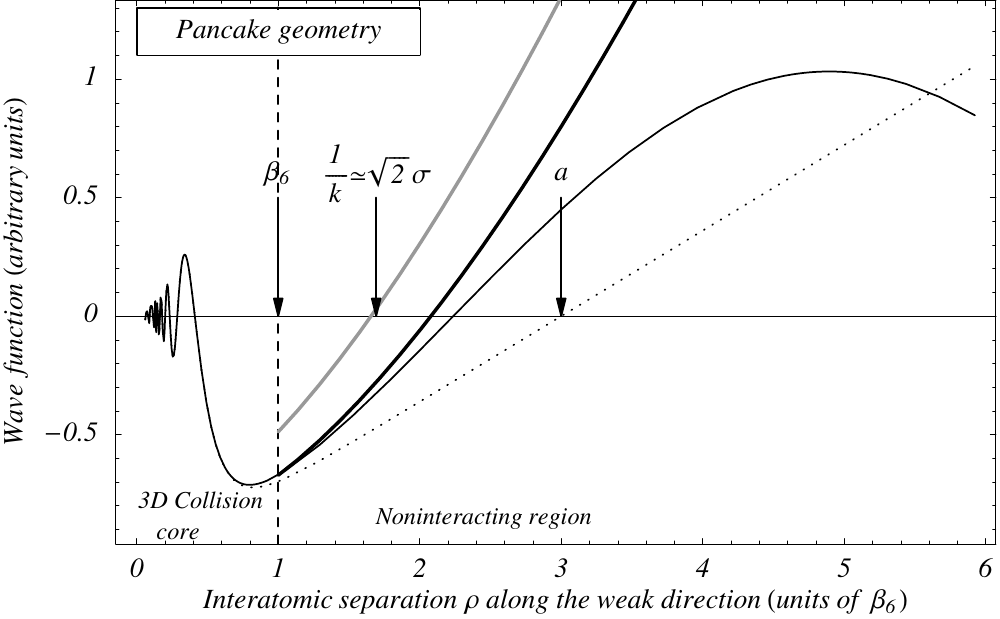}

\caption{\label{cap:Radial1}Cut through the function $r\psi(\vec{r})$ in
a confined geometry for a van der Waals interaction with scattering
length $a=3\,\beta_{6}$ , where $\beta_{6}=\left(\frac{2\mu C_{6}}{\hbar^{2}}\right)^{1/4}$
is the van der Waals length. We have chosen $a=3\beta_{6}$ as an
illustrative value; for comparison, $a\approx0.6\,\beta_{6}$ for
$^{87}\textrm{Rb}$, $a\approx-10\,\beta_{6}$ for $^{133}\textrm{Cs}$,
and $a>4\,\beta_{6}$ for $^{86}\textrm{Sr}$. \textbf{Left panel}:
function $z\psi(\rho=0,z)$ as a function of $z$ for 2D confinement
(tube) with $\sigma=1.95\,\beta_{6}$. \textbf{Right panel}: function
$\rho\psi(\rho,z=0)$ as a function of $\rho$ for 1D confinement
(pancake) with $\sigma=1.18\,\beta_{6}$. In both panels, the $s$-wave
component of the solution to the free-space scattering problem at
energy $E=\hbar^{2}k^{2}/2\mu$ is represented as a thin black line.
For $r>\beta_{6}$ it has the asymptotic form corresponding to Eq.~(\ref{eq:IntermediateRegion}).
The thick black line represents the 1D or 2D wave function in the
noninteracting region obtained from Eq.~(\ref{eq:2DLatticeWF}) or
(\ref{eq:1DLatticeWF}). It is determined by a matching procedure
with the free-space scattering wave function, as explained in the
text. The previous theories \cite{olshanii1998,petrov2000a} were
based on a matching with the solution to free-space scattering at
zero energy (dotted line), which has the asymptotic form corresponding
to Eq.~(\ref{eq:IntermediateRegion2}) for $r>\beta_{6}$. The resulting
noninteracting 1D and 2D wave functions are represented by thick grey
lines. They do not connect to the zero-energy wave function for the
considered confinements. The lengths $\beta_{6}$, $a$ and $1/k$
are indicated by arrows.}
\end{figure*}

Any scattering solution $\psi(\vec{r})$ of Eq.~(\ref{eq:Schrodinger})
is composed of an incident wave and a scattered wave. A plane wave
basis can be used for the incident wave, and the scattered wave can
be expressed with the noninteracting Green's function $G(\vec{r},\vec{r}^{\prime})$
of the system for $U(r)=0$. Namely, for 2D confinement, one has:\begin{equation}
\psi(\vec{r})\;=\;\phi_{nm}(\vec{\rho})e^{iq_{nm}z}-\int G_{2D}(\vec{r},\vec{r}^{\prime})U(r^{\prime})\psi(\vec{r}^{\prime})d^{3}\vec{r}^{\prime}\label{eq:Exact1D}\end{equation}
where $\phi_{nm}(\vec{\rho})$ denotes the unit-normalised 2D isotropic
harmonic oscillator eigenstate of principal quantum number $n$ and
angular quantum number $m$, and $q_{nm}=\sqrt{\frac{2\mu}{\hbar^{2}}E-(2n+1+\vert m\vert)/\sigma^{2}}$
is the wave number of the incident plane wave. The Green's function
reads:

\[
G_{2D}(\vec{r},\vec{r}^{\prime})=\sum_{\nu,\mu}\!\phi_{\nu\mu}(\vec{\rho})\phi_{\nu\mu}(\vec{\rho})\frac{e^{iq_{\nu\mu}\vert z-z^{\prime}\vert}}{2iq_{\nu\mu}}\]

For 1D confinement, one has:\begin{equation}
\psi(\vec{r})\;=\;\varphi_{n}(z)e^{i\vec{q}_{n}\cdot\vec{\rho}}-\int G_{1D}(\vec{r},\vec{r}^{\prime})U(r^{\prime})\psi(\vec{r}^{\prime})d^{3}\vec{r}^{\prime}\label{eq:Exact2D}\end{equation}
where $\varphi_{n}(z)$ denotes the unit-normalised 1D harmonic oscillator
eigenstate of vibrational index $n$, and $\vec{q}_{n}$ is the wave
vector of the incident plane wave with norm $q_{n}=\sqrt{\frac{2\mu}{\hbar^{2}}E-(n+\frac{1}{2})/\sigma^{2}}$
. The Green's function reads:\[
G_{1D}(\vec{r},\vec{r}^{\prime})=\sum_{\nu}\varphi_{\nu}(z)\varphi_{\nu}(z^{\prime})\frac{i}{4}H_{0}^{(1)}(q_{\nu}\vert\rho-\rho^{\prime}\vert)\]
 where $H_{\alpha}^{(1)}$ is the first Hankel function.

We first assume that the interaction potential $U(r)$ has a finite
range $r_{b}$. This means that there is a separation $r_{0}\gg r_{b}$
beyond which the wave function is essentially a solution of Eq.~(\ref{eq:Schrodinger})
with $U(r)=0$, that is to say a solution of the noninteracting problem.
This is indeed the case for typical atomic interactions which drop
off as $-C_{6}/r^{6}$van der Waals potentials. Because of this fast
drop-off, the wave function (at sufficiently low energy) reaches its
noninteracting form for $r^{4}\gg\frac{1}{20}\beta_{6}^{4}$, where
$\beta_{6}=\left(\frac{2\mu C_{6}}{\hbar^{2}}\right)^{1/4}$ \cite{julienne1996}.
The separation $r_{0}$ is therefore on the order of $\beta_{6}$,
which ranges typically from 2 nm to 5 nm.

We further assume that the interaction potential $U$ scatters only
$s$ waves. The scattering of partial waves of arbitrary order under
cylindrical confinement was treated in detail in Ref.~\cite{kim2005}.
 Retaining only $s$-wave scattering is valid in the absence of shape
resonances and if the cold-collision condition $kr_{0}<1$ is satisfied,
where $\hbar k=\sqrt{2\mu E}/\hbar$ is the collisional momentum (see
Appendix).

Under these two assumptions, the noninteracting form of Eqs~(\ref{eq:Exact1D}-\ref{eq:Exact2D})
is obtained by first taking the limit $r\gg r^{\prime}$ of the Green's
function (since $r>r_{0}\gg r_{b}\sim r^{\prime}$) and then approximating
the remaining integral over $\vec{r}^{\prime}$ by $\phi_{\nu,\mu}(0)$
(or $\varphi_{\nu}(0)$) times a quantity $4\pi A$ which does not
depend on the indices $\nu$ or $\mu$ appearing inside the Green's
funtion (taking into account the dependence on $\nu,\mu$ would introduce
higher-order partial wave scattering \cite{kim2005} - see Appendix).
These two steps are formally equivalent to taking the Green's function
out of the integral in Eqs.~(\ref{eq:Exact1D}-\ref{eq:Exact2D})
and evaluate it at $r^{\prime}=0$. Note that this can be achieved
at any $r$ by replacing the potential $U(r)$ by a regularised contact
interaction \cite{fermi1934}. For 2D confinement, one finds \cite{petrov2001}\begin{equation}
\psi(\vec{r})\;\mathop{=}_{r>r_{0}}\phi_{nm}(\vec{\rho})e^{iq_{nm}z}-4\pi A_{nm}\sum_{\nu,\mu}\!\phi_{\nu\mu}(\vec{\rho})\phi_{\nu\mu}(0)\frac{e^{iq_{\nu\mu}\vert z\vert}}{2iq_{\nu\mu}}\label{eq:2DLatticeWF}\end{equation}
and for 1D confinement \cite{olshanii1998},\begin{equation}
\psi(\vec{r})\;\mathop{=}_{r>r_{0}}\varphi_{n}(z)e^{i\vec{q}_{n}\cdot\vec{\rho}}-4\pi A_{n}\sum_{\nu}\varphi_{\nu}(z)\varphi_{\nu}(0)\frac{i}{4}H_{0}^{(1)}(q_{\nu}\rho)\label{eq:1DLatticeWF}\end{equation}
wher the factors $A_{mn}$ and $A_{n}$ are to be determined. 

The wave function $\psi$ can also be expanded in spherical partial
waves:\begin{equation}
\psi(\vec{r})=\sum_{\ell=0}^{\infty}\sum_{m=-\ell}^{\ell}\psi_{\ell,m}(r)Y_{\ell,m}(\theta,\varphi)\label{eq:PartialWaveExpansion}\end{equation}
where $Y_{\ell,m}$ are the spherical harmonics and $\theta$ the
angle between $\vec{r}$ and the $z$ axis, $\varphi$ the angle between
$\vec{\rho}$ and the $x$ axis. At short separations, the confining
potential $V(\vec{r})$ is negligible, so one can expect that within
a certain range $r_{1}$ related to the confinement length $\sigma$
(see Appendix), the wave function is close to a solution of Eq.~(\ref{eq:Schrodinger})
with $V(\vec{r})=0$, that is to say, a solution of the free-space
scattering problem. Beyond $r_{0}$, the partial waves of this solution
reach their noninteracting form which is known to be a combination
of regular and irregular spherical Bessel functions. Consistent with
the $s$-wave approximation, there is no irregular Bessel function
for $\ell\ne0$, \emph{ie} no scattered partial wave. For the $s$
wave $(\ell=0)$, we have:\begin{equation}
\psi_{00}(r)=\eta\left(\frac{\sin kr}{kr}-a(k)\frac{\cos kr}{r}\right)\quad\textrm{for }r_{0}<r<r_{1},\label{eq:IntermediateRegion}\end{equation}
where $\eta$ is a normalisation factor to be determined and $a(k)=-\tan\delta_{k}/k$
is the energy-dependent \emph{s}-wave scattering length introduced
in Refs. \cite{blume2002,bolda2002} ($\delta_{k}$ is the usual $s$-wave
phase shift, related to the $s$-wave component of the reactance matrix
$K_{s}(k)=-\tan\delta_{k}$). This energy-dependent scattering length
contains all the effects of the interaction on the wave function in
the region $r_{0}<r<r_{1}$, and for any collisional energy $E$.
For moderately tight traps $\sigma\gg r_{0}$ leading to small collisional
energies, there is a range of $r$ for which Eq.~(\ref{eq:IntermediateRegion})
simplifies to:\begin{equation}
\psi(\mathbf{r})=\eta\left(1-\frac{a}{r}\right)\label{eq:IntermediateRegion2}\end{equation}
where $a=\lim_{k\to0}a(k)$ is the scattering length of the potential.
However, for very tight lattices, $\sigma$ may be close to $r_{0}$
and only Eq.~(\ref{eq:IntermediateRegion}) holds.

\begin{figure*}[t]
\includegraphics[scale=0.85,angle=90]{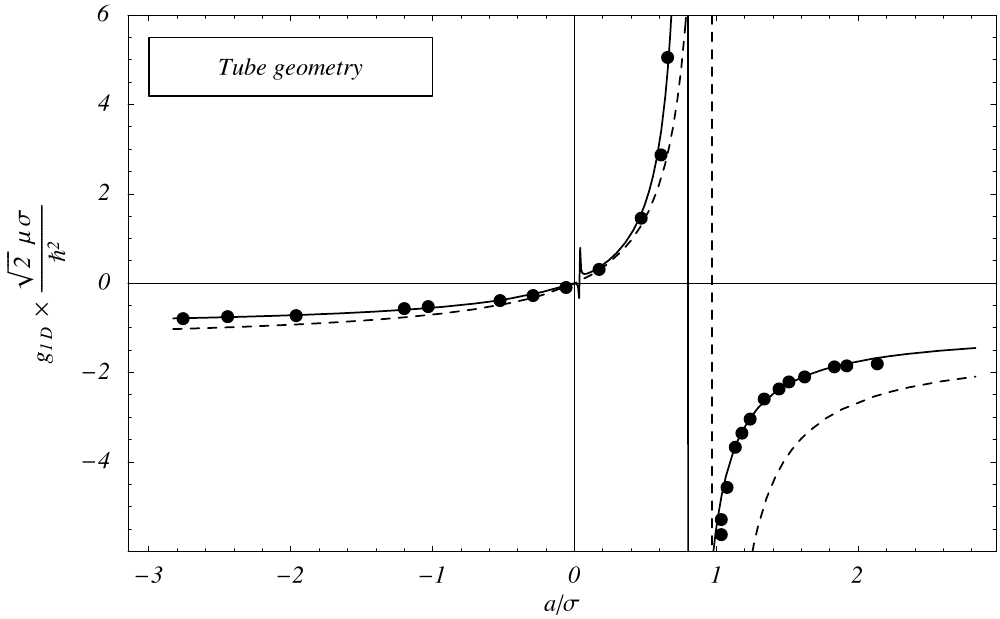}\hfill{}\includegraphics[scale=0.85]{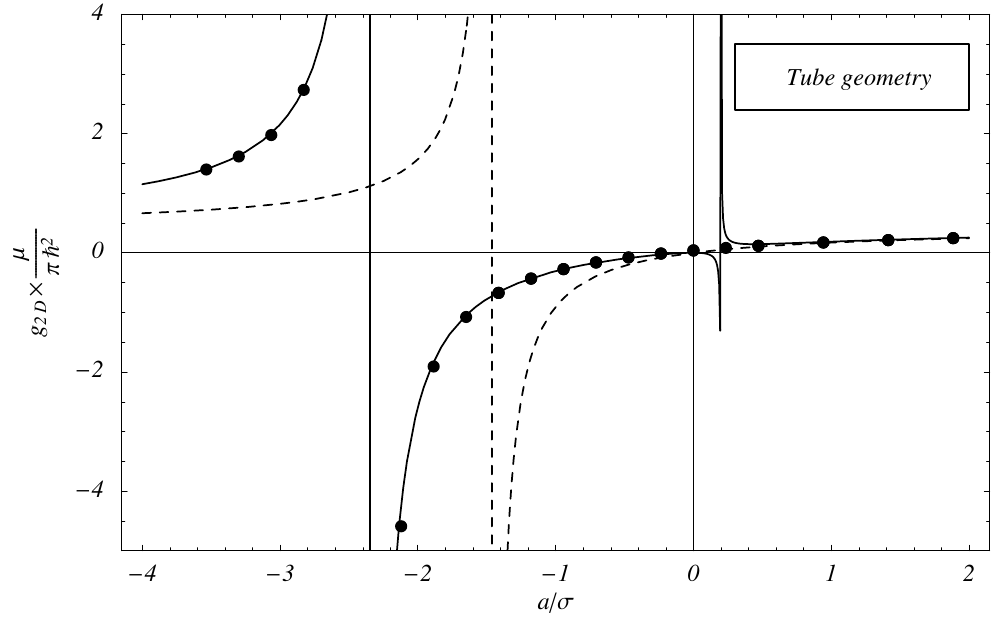}

\caption{\label{cap:g1D}The 1D interaction strength  (left panel) and the
2D interaction strength (right panel) in dimensionless units as a
function of the ratio of the 3D scattering length $a$ over the confinement
dimension $\sigma$ ($a$ is varied and $\sigma$ is fixed). As in
Fig.~\ref{cap:Radial1}, $\sigma=1.95\,\beta_{6}$ in left panel
and $\sigma=1.18\,\beta_{6}$ in the right panel. The dots are obtained
by numerically solving the problem of two atoms interacting through
a Lennard-Jones potential in a 3D cylindrically-symmetric trap; dots
in the left panel (tube geometry) were taken from Ref.~\cite{bergeman2003a}
and confirmed by us, while we calculated the dots in right panel (pancake
geometry) using the adaptive grid refinement method of Ref.~\cite{mitchell2005}.
The dashed curves correspond to the formulæ (\ref{eq:g1D}) and (\ref{eq:g2D})
with the zero-energy scattering length $a$, and the solid curves
corresponds to the same formulæ using the energy-dependent scattering
length $a(k)$ instead of $a$. Here, $a(k)$ has been calculated
in the effective-range approximation, Eq.~(\ref{eq:EffectiveRangeAk}),
which is valid almost everywhere except for small positive scattering
lengths where the approximation causes a spurious resonance.}
\end{figure*}

The essence of the method used in Refs. \cite{olshanii1998} and \cite{petrov2001}
is to assume that $\sigma\gg r_{0}$ and match the noninteracting
expressions (\ref{eq:2DLatticeWF}) and (\ref{eq:1DLatticeWF}), respectively,
with the free-space expression (\ref{eq:IntermediateRegion2}) in
the region $r_{0}<r<r_{1}$ where they are both valid. (In Ref.~\cite{olshanii1998},
this is implicitly achieved by use of a 3D regularised contact interaction).
By performing the matching procedure up to zeroth order in the asymptotic
expansion in $r/\sigma$ near $r=0$, they obtain two relations yielding
the two unknown factors $\eta$ and $A_{nm}$ (or $A_{n}$). From
that knowledge, they deduce the effective 1D and 2D interaction strengths
in the quasi-1D or 2D regime ($q_{n},q_{nm}\ll k$):\begin{eqnarray}
g_{1D} & = & \frac{\hbar^{2}}{\mu\sigma}\left(\frac{\sigma}{a}+\frac{\zeta(1/2)}{\sqrt{2}}\right)^{-1}\label{eq:g1D}\\
g_{2D} & = & 4\pi\frac{\hbar^{2}}{\mu}\left(\frac{\sqrt{2\pi}\sigma}{a}+\ln\left(\frac{B}{\pi q_{0}^{2}\sigma^{2}}\right)\right)^{-1}\label{eq:g2D}\end{eqnarray}
with $B\approx0.915$ and $\zeta(1/2)/\sqrt{2}\approx-1.033$, where
$\zeta$ is the Riemann zeta function. The singularity in these expressions
as a function $a$, $\sigma$, or $q_{0}$ corresponds to the confinement-induced
resonance. Note, however, that these analytical formulæ are valid
only when $\sigma$ is large with respect to $r_{0}$.

We stress here that the method can be extended by matching the $s$-wave
component of the noninteracting expressions (\ref{eq:2DLatticeWF})
or (\ref{eq:1DLatticeWF}) with the more general free-space expression
(\ref{eq:IntermediateRegion}). It is straightforward to see that
a matching to zeroth order in $r/\sigma$ results in the same conditions
as those of Refs.~\cite{petrov2001,olshanii1998}, but with the scattering
length $a$ being replaced by the energy-dependent $a(k)$ in all
the formulæ. In particular, $a$ is replaced by $a(k)$ in the expressions
for $\eta$, $A_{nm}$, $A_{n}$, and Eqs.~(\ref{eq:g1D}-\ref{eq:g2D}).
Surprisingly, we find that once the two unknown factors $\eta$ and
$A_{nm}$ (or $A_{n}$) are determined this way, the expression (\ref{eq:2DLatticeWF})
or (\ref{eq:1DLatticeWF}) match the expression (\ref{eq:IntermediateRegion})
up to second order in $r/\sigma$, $ie$ the next two orders automatically
match without the need for extra parameters. This was checked numerically
both in 2D and 1D confinement, and we give in the Appendix an analytical
derivation of this result for 2D confinement, based on mathematical
assumptions which we checked numerically. As a result, the expressions
(\ref{eq:2DLatticeWF}) or (\ref{eq:1DLatticeWF}) and (\ref{eq:IntermediateRegion})
can be matched within less than 1\% for $\sigma\gtrsim r_{0}$. In
Fig.~\ref{cap:Radial1} we illustrate the matching of the wave functions
for a confinement $\sigma$ close to $r_{0}$, in the case of a van
der Waals interaction $U(r)=-\frac{C_{6}}{r^{6}}$, for which $r_{0}\approx\beta_{6}$,
as stated earlier. The figure shows both the $s$-wave component
of the solution to the 3D free-space problem (which is highly oscillatory
for $r<r_{0}$, and has the asymptotic form (\ref{eq:IntermediateRegion})
for $r>r_{0}$) and the noninteracting wave function along either
the $z$ or $\rho$ directions. For better comparison, we should plot
only the $s$-wave component of the noninteracting wave function,
but this happens to be very slowly converging numerically. The inclusion
of higher-order partial waves thus induces a second-order difference
(because $\psi_{2,m}(r)\propto r^{2}$ for small $r$), but one can
still appreciate the very good matching of the two functions even
at these extreme confinements. In contrast, the two functions calculated
from the original theory do not match.

The fact that we can gain two orders in the matching procedure by
using the free-space expression (\ref{eq:IntermediateRegion}) implies
that the replacement $a\to a(k)$ greatly improves the accuracy and
the range of validity of the original theory. Although we do not have
an estimation of the accuracy on $g_{1}$ and $g_{2}$, we expect
the extended theory to give reasonnably good predictions for $\sigma\gtrsim r_{0}$.
This improvement brought by the simple replacement $a\to a(k)$ not
so surprising in view of the results reported in Refs.~\cite{bolda2002,blume2002,bolda2003,stock2005,idziaszek2006},
which have shown the relevance of energy-dependent scattering lengths
for the accurate calculation of energy levels in 3D confined geometries.
It has been long known in other contexts that contact interaction
models are significantly improved when the coupling constants are
proportional to the energy-dependent scattering length (or reactance
matrix) \cite{masnou-seeuws1982} . Similar extensions of (\ref{eq:g1D})
and (\ref{eq:g2D}) were considered in Refs.~\cite{wouter2003,granger2004,yurovsky2005}
in order to take into account the energy dependence due to a scattering
resonance at low energy. In Ref.~\cite{yurovsky2005}, a renormalised
contact interaction was used, leading to the replacement of $a$ by
the quantity $\frac{2\mu}{4\pi\hbar^{2}}T(k)$, where $T(k)=\frac{4\pi\hbar^{2}}{2\mu}a(k)(1+ika(k))^{-1}$
is the $T$-matrix. This complex quantity is equivalent to the real
$a(k)$ at low energy. Here we focus on the energy dependence for
strong confinement even in the absence of any resonance.

In the quasi-1D ($q_{nm}\ll k$) or quasi-2D ($q_{n}\ll k$) regimes,
the collisional energy is set by the confinement length $\sigma$
and can be relatively high, although the effective collisions in the
weak directions are cold. An interesting consequence of the extended
theory is that if the confinement is strong enough, the collisional
energy can probe the energy-dependence of $a(k)$. For a standard
contact potential \cite{fermi1934}, $a(k)$ is constant and equal
to $a$. However, a more realistic $a(k)$ has some energy dependence.
For instance, in the effective-range approximation, $a(k)$ has the
resonant form:\begin{equation}
a(k)\approx\frac{a}{1-\frac{1}{2}k^{2}ar_{e}},\label{eq:EffectiveRangeAk}\end{equation}
where $r_{e}$ is the effective range of the potential \cite{bethe1949}.
This approximation works well for short-range interactions with a
large scattering length $a\gg r_{e}$. In the case of van der Waals
interactions, the effective range $r_{e}$ is a simple function of
$a$ and $\beta_{6}$ \cite{flambaum1999,gao1998}:\begin{equation}
r_{e}=\frac{2}{3}\frac{\beta_{6}^{2}}{\bar{a}}\left(\left(\frac{\bar{a}}{a}\right)^{2}+\left(\frac{\bar{a}}{a}-1\right)^{2}\right)\label{eq:EffectiveRangeVdW}\end{equation}
where $\bar{a}=2\pi\beta_{6}/\Gamma(1/4)^{2}$. More elaborate analytical
expressions of $a(k)$ valid for any $a$ have been derived for van
der Waals interactions \cite{gao1998}. The interest of Eqs.~(\ref{eq:EffectiveRangeAk})
and (\ref{eq:EffectiveRangeVdW}) is that they give a simple two-parameter
description of the collisions for a wide range of energies.

To illustrate these ideas, we calculated the 1D interaction strength
$g_{1D}$ for a van der Waals interaction consistent with the Lennard-Jones
parameters of the numerical calculation reported in Ref.~\cite{bergeman2003a}.
The authors observed a difference between their numerical calculation
and the analytic formula~(\ref{eq:g1D}) where $a$ is taken as the
zero-energy scattering length. They suggested that this difference
comes the fact that the confinement-induced resonance in $g_{1D}$
results from a Feshbach resonance with a trap bound state, whose binding
energy is not predicted accurately by a pseudopotential based only
on the scattering length. As a result, the formula~(\ref{eq:g1D})
does not predict the resonance at the right location. However, we
show in Fig.~\ref{cap:g1D} that the same formula used in conjunction
with the replacement $a\to a(k)$ in the effective range approximation
reproduces the numerical calculations very well. This is because the
effective range approximation is able to reproduce the binding energy
of the last bound state accurately. The only region where the effective
range approximation fails is for small scattering lengths $a\ll\beta_{6}$,
where it predicts a spurious resonance, as visible in Fig.~\ref{cap:g1D}.

We also calculated the 2D interaction strength and checked that a
similar situation occurs in the pancake configuration. Using the adaptive
grid refinement method of Ref.~\cite{mitchell2005}, we solved the
Schr\"odinger equation (\ref{eq:Schrodinger}) for a Lennard-Jones
interaction $U(r)=\frac{C_{12}}{r^{12}}-\frac{C_{6}}{r^{6}}$ and
a cylindrical harmonic trap. The tight pancake limit is obtained by
setting the ratio of axial and radial frequencies to 400 (thus leading
to a spatial aspect ratio of 1/20), and the tight confinement scale
is set to $\sigma=1.18\,\beta_{6}$. The parameter $C_{12}$ is adjusted
to set the number of bound states supported by the interaction and
the scattering length. From this calculation, we obtained the eigenenergies
and then used Eq.~(21) of Ref.~\cite{busch1998} to extract the
2D scattering length. We found that it shows very little dependence
on the number of bound states, which can be as low as 2, saving computational
efforts. Using Eq.~(7) of Ref.~\cite{petrov2001} (or Eq.~(15)
of Ref.~\cite{lee2002}), we could then relate the 2D scattering
length to the interaction strength $g_{2D}$ for any $q_{0}$ - we
chose a $q_{0}$ given by the zero-point momentum in the weak direction.
Figure (\ref{cap:g1D}) compares this numerical $g_{2D}$ with the
analytical formula (\ref{eq:g1D}) for the same $q_{0}$. Again, the
position of the confinement-induced resonance for negative scattering
lengths \cite{petrov2000a} is correctly predicted by (\ref{eq:g1D})
provided the energy-dependent scattering length is used. As previously,
the effective range approximation works well, except for small scattering
lengths. These results also suggest that the observation of the resonance
may provide useful information about the effective range of the interaction.

In summary, we have shown that the effective 1D or 2D interactions
of ultracold bosons in strongly confined systems are governed by 3D
collisions at a relatively high energy determined by the confinement.
The effect of these high-energy collisions can be well described by
a single quantity, the energy-dependent scattering length, up to extremely
tight confinements. For van der Waals interactions, this quantity
itself can be expressed in the effective range approximation in terms
of the zero-energy scattering length and the van der Waals length.
This parametrized energy-dependent scattering length leads to an accurate
analytic prediction of the confinement-induced resonance both in 1D
and 2D confinements.

\section*{Appendix}

In this appendix, we show that for 2D confinement the matching procedure
is effective up to second order in the expansion in $r/\sigma$ near
$r=0$. Without loss of generality, we take the wave function to be
symmetric around the $z$ axis ($ie$ $m=0$ in Eq.~(\ref{eq:2DLatticeWF})),
so that Eq.~(\ref{eq:PartialWaveExpansion}) can be written as\begin{equation}
\psi(\vec{r})=\sum_{\ell=0}^{\infty}\psi_{\ell}(r)i^{\ell}(2\ell+1)P_{\ell}(\cos\theta)\label{eq:SimplifiedPartialWaves}\end{equation}
where $P_{\ell}(x)$ is the Legendre polynomial. The basic approximation
is to assume that the components $\psi_{\ell}$ are proportional to
those of the free-space solution. With the assumption that only the
$s$-wave component is scattered by the potential, we set:\begin{eqnarray}
\psi_{0}(r) & {\displaystyle \mathop{=}_{r>r_{0}}} & \eta_{0}\phi_{n0}(0)\big(j_{0}(kr)+ka(k)y_{0}(kr)\big)\label{eq:swaveComponent}\\
\psi_{\ell\ne0}(r) & {\displaystyle \mathop{=}_{r>r_{0}}} & \eta_{\ell}\phi_{n0}(0)\, j_{\ell}(kr)\label{eq:lwaveComponent}\end{eqnarray}
where $j_{\ell}$ and $y_{\ell}$ are the spherical Bessel functions,
and $\eta_{\ell}$ are factors to be determined. (Note that with this
definition, $\eta$ in Eq.~(\ref{eq:IntermediateRegion}) is equal
to $\eta_{0}\phi_{n0}(0)$). Since only components of even $\ell$
are coupled to the $s$-wave component by the trapping potential $V(\vec{r})$,
we can discard odd-$\ell$ components (they do not play a role in
the scattering process). An expansion near $r=0$ of Eq.~(\ref{eq:SimplifiedPartialWaves})
therefore reads\begin{widetext}\begin{equation}
\phi_{n0}(0)\left[\eta_{0}\left(-\frac{ka(k)}{kr}+1\right)+\left(\eta_{0}\frac{ka(k)}{2}\right)kr-\left(\frac{\eta_{0}}{6}+\frac{\eta_{2}}{3}P_{2}(\cos\theta)\right)(kr)^{2}-\left(\eta_{0}\frac{ka(k)}{24}\right)(kr)^{3}+\dots\right]\label{eq:1Dapproximate}\end{equation}
\end{widetext}This is to be matched with Eq.~(\ref{eq:2DLatticeWF}).
Using the explicit form of the harmonic oscillator solutions $\phi_{n0}$,
it can be written as

\begin{equation}
\phi_{n0}(0)\Upsilon_{n}\left[\frac{\rho}{\sqrt{2}\sigma},\,\frac{z}{\sqrt{2}\sigma},\,-\frac{1}{2}q_{00}^{2}\sigma^{2}\right]\label{eq:1DnoninteractingUnitless}\end{equation}
with \begin{eqnarray*}
\Upsilon_{n}\left[u,\, v,\,\epsilon\right] & = & e^{-\frac{u^{2}}{2}}L_{n}(u^{2})\cosh(2v\sqrt{n+\epsilon})-\frac{A_{n0}}{\sqrt{2}\sigma}\Lambda\left[u,v,\epsilon\right]\\
\Lambda\left[u,\, v,\,\epsilon\right] & = & \sum_{\nu=0}^{\infty}\frac{e^{-2v\sqrt{\nu+\epsilon}}}{\sqrt{\nu+\epsilon}}e^{-\frac{u^{2}}{2}}L_{\nu}(v^{2})\end{eqnarray*}
where the $L_{\nu}$ is the Laguerre polynomial. For Eq.~(\ref{eq:1DnoninteractingUnitless})
a partial wave expansion appears not feasible, so we resort to evaluating
the expressions along the $\rho$ and $z$ axes. Along the $z$ axis
($\rho=0$), we find the following expansion\begin{eqnarray}
\Lambda[0,v,\epsilon] & = & \frac{1}{v}+\zeta(\!\!\begin{array}{c}
\frac{1}{2}\!\end{array},\epsilon)+(2\epsilon-1)v+2\zeta(\!-\begin{array}{c}
\!\frac{1}{2}\end{array},\epsilon)v^{2}\nonumber \\
 &  & +\frac{1}{6}\left[(2\epsilon-1)^{2}-\frac{1}{3}\right]v^{3}+\dots\label{eq:lambda1}\end{eqnarray}
Along the $\rho$ axis ($z=0$), we find the following expansion\begin{eqnarray}
\Lambda[u,0,\epsilon] & = & \frac{1}{u}+\zeta(\begin{array}{c}
\!\!\frac{1}{2}\!\end{array},\epsilon)+(2\epsilon-1)u\nonumber \\
 &  & +\left[\frac{1}{2}(2\epsilon-1)\zeta(\begin{array}{c}
\!\!\frac{1}{2}\!\end{array},\epsilon)-\zeta(\!-\begin{array}{c}
\!\frac{1}{2}\end{array},\epsilon)\right]u^{2}\nonumber \\
 &  & +\frac{1}{6}\left[(2\epsilon-1)^{2}+\frac{2}{3}\right]u^{3}+\dots\label{eq:lambda2}\end{eqnarray}
In these expansions, $\zeta$ refers to the Hurwitz zeta function
which we define as follows\[
\zeta(s,\epsilon)=\lim_{N\to\infty}\sum_{n=0}^{N}\frac{1}{(n+\epsilon)^{s}}-\frac{(N+\epsilon+\frac{1}{2})^{-s+1}}{-s+1}\]
for $-1<s<1$, generalizing the definition given in Refs.~\cite{bergeman2003a,moore2004}.
The first expansion (\ref{eq:lambda1}) was found using the counter-term
method explained in these references with the refined counter term
$\int_{0}^{N+\frac{1}{2}+\epsilon}\frac{\exp(-2\sqrt{s}x)}{\sqrt{s}}ds$.
The second expansion (\ref{eq:lambda2}) was guessed from the expected
result (\ref{eq:1Dapproximate}), and checked numerically. Because
of the very slow convergence of the sum in $\Lambda[u,0,\epsilon]$
, we could not check the terms directly. Instead, we noted that the
Laplace transform of $\Lambda$ with respect to the argument $u$
is related to the Lerch transcendent $\Phi$ \cite{erdelyi1953}:\[
\int_{0}^{\infty}e^{-su}\Lambda[u,0,\epsilon]du=\frac{1}{s+\frac{1}{2}}\Phi\left(\frac{s-\frac{1}{2}}{s+\frac{1}{2}},\frac{1}{2},\epsilon\right)\]
 and assumed there is a direct correspondence between the Laplace
transform of the terms in (\ref{eq:lambda2}) and the terms in the
asymptotic expansion of the Laplace transform near $s\to\infty$.

Matching the expressions (\ref{eq:1Dapproximate}) and (\ref{eq:1DnoninteractingUnitless})
along the $z$ direction leads to the following relations for each
order of the expansion:\begin{eqnarray}
\eta_{0}a(k) & = & A_{n0}\label{eq:M1}\\
\eta_{0} & = & 1-\frac{A_{n0}}{\sqrt{2}\sigma}\zeta(\begin{array}{c}
\!\!\frac{1}{2}\!\end{array},\epsilon)\label{eq:M2}\\
\eta_{0}\frac{k^{2}a(k)}{2} & = & \frac{A_{n0}}{2\sigma^{2}}(k^{2}\sigma^{2})\label{eq:M3}\\
\left(\frac{\eta_{0}}{6}+\frac{\eta_{2}}{3}\right)\!\! k^{2} & = & \!\frac{k^{2}}{2}-\frac{1+2n}{2\sigma^{2}}+\!\frac{A_{n0}}{\sqrt{2}\sigma^{3}}\zeta(\begin{array}{c}
\!\!-\frac{1}{2}\!\end{array},\epsilon)\label{eq:M4}\\
\eta_{0}\frac{ka(k)}{24}k^{3} & = & \frac{A_{n0}}{24\sigma^{4}}\left(k^{4}\sigma^{4}-\frac{1}{3}\right)\label{eq:M5}\end{eqnarray}
while, matching along the $\rho$ direction leads to the following
relations\begin{eqnarray}
\eta_{0}a(k) & = & A_{n0}\label{eq:N1}\\
\eta_{0} & = & 1-\frac{A_{n0}}{\sqrt{2}\sigma}\zeta(\begin{array}{c}
\!\!\frac{1}{2}\!\end{array},\epsilon)\label{eq:N2}\\
\eta_{0}\frac{k^{2}a(k)}{2} & = & -\frac{A_{n0}}{2\sigma^{2}}(k^{2}\sigma^{2})\label{eq:N3}\\
\left(\frac{\eta_{0}}{6}-\frac{\eta_{2}}{6}\right)\!\! k^{2} & = & \frac{1+2n}{4\sigma^{2}}\label{eq:N4}\\
 &  & -\frac{A_{n0}}{2\sqrt{2}\sigma^{3}}\left[\frac{k^{2}\sigma^{2}}{2}\zeta(\begin{array}{c}
\!\!\frac{1}{2}\!\end{array},\epsilon)+\zeta(\!-\begin{array}{c}
\!\frac{1}{2}\end{array},\epsilon)\right]\nonumber \\
\eta_{0}\frac{ka(k)}{24}k^{3} & = & \frac{A_{n0}}{24\sigma^{4}}\left(k^{4}\sigma^{4}+\frac{2}{3}\right)\label{eq:N5}\end{eqnarray}
where we make use of $2\epsilon-1=-k^{2}\sigma^{2}$. Relations (\ref{eq:M1}-\ref{eq:M2})
and (\ref{eq:N1}-\ref{eq:N2}) are the same and determine $A_{n0}$
and the $s$-wave factor:\begin{equation}
\eta_{0}=\frac{1}{1+\frac{a(k)}{\sqrt{2}\sigma}\zeta(\begin{array}{c}
\!\!\frac{1}{2}\!\end{array},-\frac{q_{00}^{2}\sigma^{2}}{2})}\label{eq:eta0}\end{equation}

Relations (\ref{eq:M3}) and (\ref{eq:N3}) are the same and are both
satisfied with the previous determination of $A_{n0}$ and $\eta_{0}$.
Relations (\ref{eq:M4}) and (\ref{eq:N4}) are consistent and give
the same determination of the $d$-wave factor:\begin{equation}
\eta_{2}=\eta_{0}\left(\frac{3a(k)}{\sqrt{2}k^{2}\sigma^{3}}\zeta(\!-\begin{array}{c}
\!\frac{1}{2}\end{array},-\!\!\begin{array}{c}
\frac{q_{00}^{2}\sigma^{2}}{2}\end{array}\!\!)-\frac{1}{2}\right)+\frac{3}{2}\frac{q_{n0}^{2}}{k^{2}}\label{eq:eta2}\end{equation}

However, neither relation (\ref{eq:M5}) or (\ref{eq:N5}) are satisfied
with the previous determination of $A_{n0}$ and $\eta_{0}$ (because
of the terms $-\frac{1}{3}$ and $\frac{2}{3}$), which means that
the free-space approximation (\ref{eq:swaveComponent}-\ref{eq:lwaveComponent})
is valid up to order $r^{2}$. The error is $\phi_{n0}(0)\vert\eta_{0}a(k)\vert\frac{1}{24\sigma^{4}}r^{3}$.
The range of $r$ for which this error is negligible determines the
range $[r_{0},r_{1}]$ where a matching is possible. Since the error
increases with $r$, we can define $r_{1}$ as the $r$ for which
the ratio between the error and the wave function (\ref{eq:1Dapproximate})
is equal to a certain tolerance $\alpha$. In the limit of large scattering
lengths, one finds \[
r_{1}=(24\alpha)^{1/4}\sigma.\]

The range {[}$r_{0},r_{1}$] where the two functions can be matched
to within the tolerance $\alpha$ exists as long as $r_{1}>r_{0}$,
\emph{ie}\begin{equation}
\sigma>(24\alpha)^{-1/4}r_{0}.\label{eq:Condition}\end{equation}

For instance, for $\alpha=1\%$, one gets $\sigma>1.43r_{0}.$ This
indicates that the method works even for a confinement length $\sigma$
on the order of the range $r_{0}$.

We can also check that this is consistent with our assumption that
only $s$ waves are scattered. Higher-order partial wave scattering
arises if we take into account the dependence of $A_{n0}$ on $\nu$.
The first correction is $A_{n0}\to A_{n0}+(\nu+\frac{1}{2})B_{n0}$,
and leads to the term $B_{n0}\sigma^{2}/z^{3}$ along the $z$ axis.
This term is to be matched with the leading-order term of the scattered
$d$-wave term $(-K_{2})\eta_{2}\phi_{n0}(0)y_{2}(kz)(-5)P_{2}(\cos0)$,
where $K_{2}$ is the $d$-wave reactance matrix element. In the absence
of shape resonance, $K_{2}$ is purely determined by the long-range
van der Waals behaviour of the interaction, $K_{2}\approx-\frac{1}{100}(k\beta_{6})^{4}$
- see Eq.~(8) of Ref.~\cite{gao1998}. This leads to:\[
B_{n0}=-\eta_{2}\frac{\sigma}{20}\frac{(k\beta_{6})^{4}}{(k\sigma)^{3}}\]

$d$-wave scattering is negligible if $\vert B_{n0}\vert\ll\vert A_{n0}\vert$.
In the quasi-1D regime ($q_{n0}\ll k$ and $k\sim1/\sigma$), and
using the value of $\eta_{2}$ (\ref{eq:eta2}), one finds:\[
\sigma^{4}\gg\left(\frac{1}{40}\vert1-\sigma/a(k)\vert\right)\beta_{6}^{4},\]
which for a wide range of scattering lengths is consistent with the
condition (\ref{eq:Condition}) and the cold collision condition $kr_{0}<1$
given in the text.

\bibliographystyle{apsrev}
\bibliography{/home/pascal/Redaction/biblio,/home/pascal/Redaction/biblio_extra}

\end{document}